\documentclass[aps,prb,twocolumn,preprintnumbers,amsmath,amssymb,superscriptaddress,showpacs]{revtex4-1}
\pdfoutput=1
\usepackage{graphicx}
\usepackage{bm}
\usepackage{color}
\usepackage{hyperref}

\newcommand{\ket}[1]{|#1\rangle}

\newcommand{\expect}[1]{\langle #1 \rangle}
\newcommand{\llangle}{\langle\!\langle}
\newcommand{\rrangle}{\rangle\!\rangle}
\newcommand{\e}{\varepsilon}

\usepackage[normalem]{ulem}

\begin{document}

\title{Underscreened Kondo effect in $S=1$ magnetic quantum dots:\\Exchange, anisotropy and temperature effects}

\author{Maciej Misiorny}
 \email{misiorny@amu.edu.pl}
\affiliation{Peter Gr{\"u}nberg Institut, Forschungszentrum J{\"u}lich, 52425 J{\"u}lich,  Germany}
\affiliation{JARA\,--\,Fundamentals of Future Information Technology, 52425 J{\"u}lich,  Germany}
\affiliation{Faculty of Physics, Adam Mickiewicz University, 61-614 Pozna\'{n}, Poland}

\author{Ireneusz Weymann}
\affiliation{Faculty of Physics, Adam Mickiewicz University,
61-614 Pozna\'{n}, Poland}

\author{J\'{o}zef Barna\'{s}}
\affiliation{Faculty of Physics, Adam
Mickiewicz University, 61-614 Pozna\'{n}, Poland}
\affiliation{Institute of Molecular Physics, Polish Academy of
Sciences, 60-179 Pozna\'{n}, Poland
}%


\begin{abstract}
We present a theoretical analysis of the effects of uniaxial
magnetic anisotropy and contact-induced exchange field on the
underscreened Kondo effect in $S=1$ magnetic quantum dots coupled
to ferromagnetic leads. First, by using the second-order
perturbation theory we show that the coupling to spin-polarized
electrode results in an effective exchange field $B_{\rm eff}$ and
an effective magnetic anisotropy $D_{\rm eff}$. Second, we confirm
these findings by using the numerical renormalization group
method, which is employed to study the dependence of the quantum
dot spectral functions, as well as quantum dot spin, on various
parameters of the system. We show that the underscreened Kondo effect is
generally suppressed due to the presence of effective exchange
field and can be restored by tuning the anisotropy constant, when
$|D_{\rm eff}| = |B_{\rm eff}|$. The Kondo effect can also be
restored by sweeping an external magnetic field, and the
restoration occurs twice in a single sweep. From the distance
between the restored Kondo resonances one can extract the
information about both the exchange field and the effective anisotropy.
Finally, we calculate the temperature dependence of linear
conductance for the parameters where the Kondo effect is restored
and show that the restored Kondo resonances display a universal scaling of
$S=1/2$ Kondo effect.
\end{abstract}



\maketitle


\section{Introduction}

Although manifestation of the Kondo effect in nanoscopic systems
of spin $S>1/2$ has been the subject of extensive experimental and
theoretical studies for more than a
decade,~\cite{Scott_ACSNano4/2010,Sasaki_Nature405/2000,Schmid_Phys.Rev.Lett.84/2000,Wiel_Phys.Rev.Lett.88/2002,Kogan_Phys.Rev.B67/2003}
it is still attracting considerable attention. From the
experimental point of view, this was triggered by a rapid
development of
techniques~\cite{Meier_Science320/2008,Ternes_J.Phys.:Condens.Matter21/2009,Wiebe_J.Phys.D:Appl.Phys.44/2011}
allowing for controlled preparation and investigation of single
magnetic impurities, such as atoms and molecules, placed on a
surface~\cite{Zhao_Science309/2005,Voss_Phys.Rev.B78/2008,Mannini_Nature468/2010,Pruser_NaturePhys.7/2011,Kahle_NanoLett.12/2011}
or captured in a
junction.~\cite{Heersche_Phys.Rev.Lett.96/2006,Jo_NanoLett.6/2006,Roch_Nature453/2008,Zyazin_NanoLett.10/2010,
Florens_J.Phys.:Condens.Matter23/2011,Vincent_Nature488/2012,Burzuri_Phys.Rev.Lett.109/2012}
Furthermore, an important issue is the interaction of individual
large-spin atoms or molecules with the environment, which  may
contribute to a magnetic
anisotropy.\cite{Boca_book,Gambardella_Science300/2003,Hirjibehedin_Science317/2007,Wende_NatureMater.6/2007,
Brune_Surf.Sci.603/2009,Gambardella_NatureMater.8/2009,Serrate_NatureNanotech.5/2010}
A significant uniaxial magnetic anisotropy, in turn, results in an
energy barrier for switching the molecules's spin between two
metastable states -- the feature indispensable for potential
applications in information storage
technologies.~\cite{Mannini_NatureMater.8/2009,Loth_NaturePhys.6/2010}
Interestingly enough, the magnetic state of such a system can in
principle be controlled by means of spin-polarized
currents,~\cite{Timm_Phys.Rev.B73/2006,
Misiorny_Phys.Rev.B75/2007,Misiorny_Phys.Stat.Sol.B246/2009,Misiorny_Phys.Rev.B79/2009}
which has already been experimentally
confirmed.~\cite{Loth_NaturePhys.6/2010}

In order to be able to exploit advantageous features stemming from
the presence of magnetic anisotropy, a possibility of its external
control would be very desirable. Indeed, several experiments have
so far confirmed the feasibility of such a control. The most
straightforward way to modify the magnetic anisotropy of an adatom
is just to change its nearest atomic environment, which can be
achieved simply by deposition of the adatom at topologically
different points of a
substrate.~\cite{Brune_Surf.Sci.603/2009,Hirjibehedin_Science312/2006,Otte_NaturePhys.4/2008}
More elaborate techniques demonstrated for molecules involve the
application of electric field,~\cite{Zyazin_NanoLett.10/2010,Burzuri_Phys.Rev.Lett.109/2012} or
even the mechanical modification of the molecular
symmetry.~\cite{Parks_Science328/2010} In fact, the latter method
allows for a fully-controllable and continuous tuning of the
anisotropy constant, which was demonstrated for a spin $S=1$
quantum dot in the underscreened Kondo regime.
~\cite{Nozieres1980,LeHur_PhysRevB.56.668,Coleman_PhysRevB.68.220405,
Posazhennikova_PhysRevLett.94.036802,Koller_PhysRevB.72.045117,Mehta_Phys.Rev.B72/2005,Cornaglia_Europhys.Lett.93/2011}

Screening of a quantum dot spin appears when the dot becomes
strongly coupled to electrodes. For temperatures $T$ smaller than
the Kondo temperature $T_\textrm{K}$, the spin exchange processes
due to electronic correlations can lead to an additional sharp
peak in the density of states --- the Kondo-Abrikosov-Suhl
resonance. Generally, in order to observe full screening of a
magnetic impurity spin $S$, the impurity should be coupled to $2S$
screening channels.~\cite{Nozieres1980,Hewson_book,Zitko_Phys.Rev.B78/2008} In
turn, a typical experimental setup for measuring transport through
quantum dots or molecules involves usually two contacts. This
implies that when connecting the spin $S=1$ dot to two (say first
and second) leads, the spin could be in principle fully screened.~\cite{Florens_J.Phys.:Condens.Matter23/2011}
In order to observe the underscreened Kondo effect, one needs to
use a more specific setup, as demonstrated by Roch {\em et
al.}~\cite{Roch_Phys.Rev.Lett.103/2009} Since the screening
becomes effective when $T<T_\textrm{K}$~\cite{Ferry_book} and the
Kondo temperature depends exponentially on the dot-lead coupling
strength $\Gamma$, by connecting the dot asymmetrically to
external leads one obtains two different Kondo temperatures:
$T_\textrm{K}^{1(2)}$ for the first (second) lead. The
underscreened Kondo effect can be then observed when the condition
$T_\textrm{K}^1 \ll T \ll T_ {K}^2$
is fulfilled.~\cite{Roch_Phys.Rev.Lett.103/2009} In such a
case, the spin is only partially screened by electrons of the
strongly-coupled lead, while the other lead serves as a weakly
coupled probe. Despite its theoretical simplicity, the first
experimental realization of the underscreened Kondo effect was
reported only very
recently.~\cite{Roch_Phys.Rev.Lett.103/2009,Parks_Science328/2010}

In this paper, motivated e.g. by the experiments of Parks \emph{et
al.},~\cite{Parks_Science328/2010} we analyze the transport
properties of a spin $S=1$ system strongly coupled to a
ferromagnetic reservoir. In particular, we focus on discussing how
the uniaxial magnetic anisotropy and the
ferromagnetic-contact-induced exchange field affect the
underscreened Kondo effect. Our analysis is based on the {\em
full} density-matrix numerical renormalization group (fDM-NRG)
method,~\cite{Wilson_Rev.Mod.Phys.47/1975,Bulla_Rev.Mod.Phys.80/2008,
Weichselbaum_Phys.Rev.Lett.99/2007,Toth_Phys.Rev.B78/2008}
which is known as the most powerful and exact in addressing
transport properties of various nanostructures in the Kondo
regime. As we are mainly interested in the aspects of the
underscreened Kondo effect, which are related to the coexistence
of magnetic anisotropy and ferromagnetism of the screening
channel, we assume a model in which only one electrode is
attached to the dot, see Fig.~\ref{Fig:1}. Such a setup defines a
typical one-channel Kondo
experiment,~\cite{Roch_Phys.Rev.Lett.103/2009,Parks_Science328/2010}
where the role of the second (weakly coupled) electrode in the
formation of the Kondo resonance can be neglected (the
corresponding Kondo temperature tends to zero). Nevertheless, the
second electrode, being a weakly coupled probe (e.g. a tip of
STM), will enable the measurements of the conductance through the
system and the local density of states.

The paper is organized as follows. In Sec. II we describe the
model and method used in calculations. Section III is devoted to
basic concepts, where we discuss the spectrum of an isolated dot
and the effects of renormalization due to the coupling to
electrode. Numerical results and their discussion are presented in
Sec. IV with a special focus on the effects due to magnetic
anisotropy. Finally, the conclusions can be found in Sec. V.

\section{Theoretical description}
\subsection{Model}

\begin{figure}[t]
  \includegraphics[width=0.5\columnwidth]{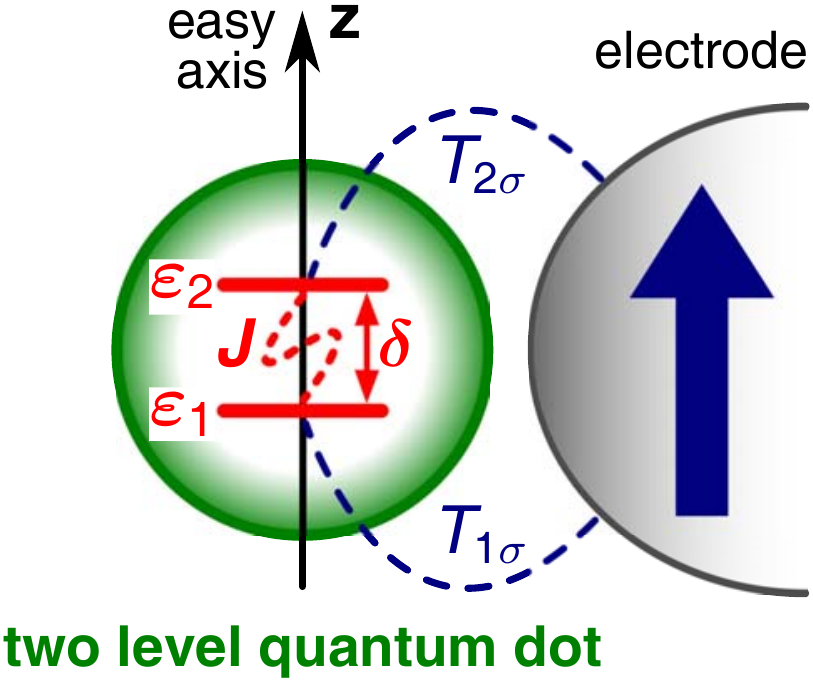}
  \caption{\label{Fig:1} (Color online)
  Schematic of a two-level magnetic quantum dot coupled to a metallic ferromagnetic electrode.
  The magnetic moment of the electrode (denoted by a bold arrow)
  is collinear with the dot's easy axis. The dot levels have energies
  $\varepsilon_1$ and $\varepsilon_2$, respectively, with $\delta$ being the level
  spacing,
  while $J$ denotes the exchange interaction.
  The tunnel matrix elements between the dot and the lead are denoted
  by $T_{j\sigma}$ for the dot level $j$ and spin $\sigma$.}
\end{figure}

The total Hamiltonian of a two-level magnetic quantum dot coupled to an external lead, see Fig.~\ref{Fig:1}, can be written as
    \begin{equation}\label{Eq:1}
    \mathcal{H} = \mathcal{H}_\textrm{mol} + \mathcal{H}_\textrm{lead} + \mathcal{H}_\textrm{tun},
    \end{equation}
where the first term describes the quantum dot, the second one
refers to the lead, whereas the final term represents tunneling
processes between the dot and the lead.

A bare two-level quantum dot can be characterized by the model Hamiltonian
    \begin{align}\label{Eq:2}
    \mathcal{H}_\textrm{mol} =& \sum_{j=1,2}\sum_{\sigma=\uparrow,\downarrow}\!\! \varepsilon_j\,n_{j\sigma}  +
    U\!\! \sum_{j=1,2}\!\! n_{j\uparrow}n_{j\downarrow}  +
    U^\prime\!\!\!\!\!  \sum_{\sigma,\sigma^\prime=\uparrow,\downarrow}\!\!\!\! n_{1\sigma}n_{2\sigma^\prime}
    \nonumber\\
    & + J S^2 + DS_z^2 + B_zS_z .
    \end{align}
In the above, $n_{j\sigma}=c_{j\sigma}^\dagger c_{j\sigma}^{}$ and
$c_{j\sigma}^\dagger  (c_{j\sigma}^{})$ denotes the creation
(annihilation) operator of an electron with spin $\sigma$ and
energy $\varepsilon_j$ in the $j$th level ($j=1,2$). For
convenience, we write the energy levels as
$\varepsilon_1=\varepsilon-\delta/2$ and
$\varepsilon_2=\varepsilon+\delta/2$, where $\varepsilon$ is the
average value of the two levels while $\delta$ is the level
spacing. The Coulomb energy of two electrons of opposite spins
occupying the same level is given by $U$ (assumed to be the same
for both levels), whereas the inter-level Coulomb correlations are
described by $U^\prime$. For simplicity, we assume equal
correlation energies $U=U'$ in the following. Furthermore,  $J$
stands for the interlevel exchange interaction with
$\bm{S}=\bm{S}_1+\bm{S}_2$, where $\bm{S}_j$ is the electron spin
operator for the dot level $j$,
$\textbf{S}_j=\frac{1}{2}\sum_{\sigma\sigma'}c_{j\sigma}^\dag
\bm{\sigma}_{\sigma\sigma'}^{} c_{j\sigma^\prime}^{}$, with
$\bm{\sigma}=(\sigma^x,\sigma^y,\sigma^z)$ being the Pauli spin
operator. According to the Hund's rules, this interaction should
be generally of a \emph{ferromagnetic} type $(J<0)$, nonetheless
the possibility of a weakly \emph{antiferromagnetic} coupling has
also been reported.~\cite{Logan_Phys.Rev.B80/2009} Finally, the
lowest order uniaxial magnetic anisotropy is represented by the
anisotropy constant $D$, and the last term of Eq.~(\ref{Eq:2})
describes the Zeeman energy of the dot in an external magnetic
field $B_z$ applied along the dot's easy axis, with $g\mu_\textrm{B}\equiv1$.

The Hamiltonian for a ferromagnetic metallic reservoir of noninteracting itinerant electrons is given by
    \begin{equation}\label{Eq:3}
    \mathcal{H}_{\textrm{lead}}=\sum_{\bm{k}}\sum_{\sigma=\uparrow,\downarrow}\varepsilon_{\bm{k}\sigma}
    a_{\bm{k}\sigma}^{\dag} a_{\bm{k}\sigma}^{\mbox{}}.
    \end{equation}
Here, $a_{\bm{k}\sigma}^{\dag}( a_{\bm{k}\sigma}^{\mbox{}})$
creates (annihilates) an electron of energy
$\varepsilon_{\bm{k}\sigma}$, where $\bm{k}$ indicates a wave
vector, while $\sigma$ is a spin index of the electron. It is
important to mention that in the following discussion we assume
that magnetic moment of the electrode remains collinear to the
dot's easy axis.

Finally, tunneling of electrons between the electrode and the dot is described in general by
    \begin{equation}\label{Eq:4}
    \mathcal{H}_\textrm{tun} = \sum_{\bm{k}}\sum_{j=1,2}\sum_{\sigma=\uparrow,\downarrow}
    T_{j\bm{k}\sigma}^{} a_{\bm{k}\sigma}^{\dag} c_{j\sigma^{}} + \textrm{H.c.},
    \end{equation}
where $T_{j\bm{k}\sigma}$ denotes the tunnel matrix element
between the dot's $j$th level and the  electrode, see
Fig.~\ref{Fig:1}. In the following we assume that both levels are
coupled symmetrically to the electrode, i.e. $T_{1\bm{k}\sigma} =
T_{2\bm{k}\sigma}\equiv T_{\bm{k}\sigma}$. Although such
foundation is not the most general
one,~\cite{Pustilnik_Phys.Rev.Lett.87/2001,Posazhennikova_Phys.Rev.B75/2007}
it is sufficient for the present analysis of the effects resulting
from magnetic anisotropy and exchange field in the context of the
underscreened Kondo problem. In order to further facilitate
calculations, we assume that the full spin-dependence is included
exclusively \emph{via} the matrix elements
$T_{\sigma}$,~\cite{Choi_Phys.Rev.Lett.92/2004,Sindel_Phys.Rev.B76/2007}
where the $\bm{k}$-dependence has also been neglected. In
addition, we assume a symmetric and flat conduction band extending
within the range $[-W,W]$, so that the density of states is
$\rho(\omega)\equiv\rho=1/(2W)$, and we use $W\equiv1$ as the
energy unit. Consequently, the spin-dependent hybridization
function reads, $\Gamma_\sigma=\pi\rho |T_\sigma|^2$. Now,
introducing the spin polarization coefficient $P$ of the
electrode, defined as
$P=(\Gamma_\uparrow-\Gamma_\downarrow)/(\Gamma_\uparrow+\Gamma_\downarrow)$,
the spin-dependent coupling can be parameterized as:
$\Gamma_{\uparrow(\downarrow)}=\Gamma(1\pm P)$, with
$\Gamma=(\Gamma_\uparrow+\Gamma_\downarrow)/2$.

\subsection{Objectives and method of calculations}

The main quantity we are interested in is the zero-temperature,
spin-dependent equilibrium \emph{spectral function} of the quantum dot
    \begin{equation}\label{Eq:6}
    A_\sigma^{jj^\prime}(\omega)=-\frac{1}{\pi} \Im{\rm m} \llangle c_{j\sigma}^{}|c_{j^\prime\sigma}^{\dagger}\rrangle_\omega^\textrm{r}\ \ \ (j,j^\prime=1,2),
    \end{equation}
with $\llangle
c_{j\sigma}^{}|c_{j^\prime\sigma}^{\dagger}\rrangle_\omega^\textrm{r}$
standing for the Fourier transform of the retarded Green's
function $\llangle
c_{j\sigma}^{}|c_{j^\prime\sigma}^{\dagger}\rrangle_t^\textrm{r}
=-i\theta(t)\langle\{c_{j\sigma}^{}(t),c_{j^\prime\sigma}^{\dagger}(0)\}\rangle$.
It is worth noting that the spectral function with two identical
indices, i.e. $A_\sigma^{jj}(\omega)$, is related to the
spin-resolved density of states associated with  the $j$th level,
whereas $A_\sigma^{jj^\prime}(\omega)$ with $j\neq j^\prime$
corresponds to processes of electrons entering and leaving the dot
at different levels. Because measuring the spin-resolved
components of the spectral function may pose a serious
experimental challenge, we will focus on discussing the total
spectral function. Thus, we introduce the normalized full spectral
function $A(\omega)$,
    \begin{equation}\label{Eq:7}
    A(\omega)=\pi \sum_{jj^\prime}\sum_\sigma \Gamma_\sigma
    A_\sigma^{jj^\prime}(\omega).
    \end{equation}
The importance of the spectral function $A(\omega)$ stems from the fact that in a
two-terminal setup with the second lead being a weakly-coupled
probe, e.g. a tip of an STM microscope, the differential
conductance of the system at bias voltage $eV$ can be approximated
as
$\tfrac{\textrm{d}I}{\textrm{d}V}\sim\tfrac{e^2}{h}A(\omega=eV)$.~\cite{Bruus_book,Csonka2012}
On the other hand,  the spectral function for $\omega\to 0$, $A(0)$,  determines the
linear-response conductance.

In the light of the preceding discussion, the central problem is
the calculation of the spectral function $A_\sigma^{jj^\prime}(\omega)$. In
the Kondo regime, this can be reliably done by means of the numerical
renormalization group
(NRG),~\cite{Hewson_book,Wilson_Rev.Mod.Phys.47/1975,Bulla_Rev.Mod.Phys.80/2008}
which enables us to analyze the static and dynamic properties of the
system in the most accurate manner. The essential idea of the
method lies in a logarithmic discretization of the conduction
band and mapping of the system's Hamiltonian onto a semi-infinite
chain, with the quantum dot residing at the initial site.
Iterative diagonalization of the Hamiltonian by adding consecutive
sites of the chain allows then for resolving key properties of the
system at energy scale $\Lambda^{-n/2}$, with $\Lambda>1$ denoting
the discretization parameter and $n$ a given iteration.

In order to address the present problem efficiently,
the calculations have been performed with the use of the
flexible density-matrix numerical renormalization group
(DM-NRG) code.~\cite{Toth_Phys.Rev.B78/2008,Legeza_DMNRGmanual}
In this study we exploited the $U_\textrm{charge}(1)\times U_\textrm{spin}(1)$ symmetries
corresponding to conservation of the electron number (charge) and
the $z$th component of the total spin.

\section{Basic concepts}

\begin{figure}[t]
  \includegraphics[width=0.85\columnwidth]{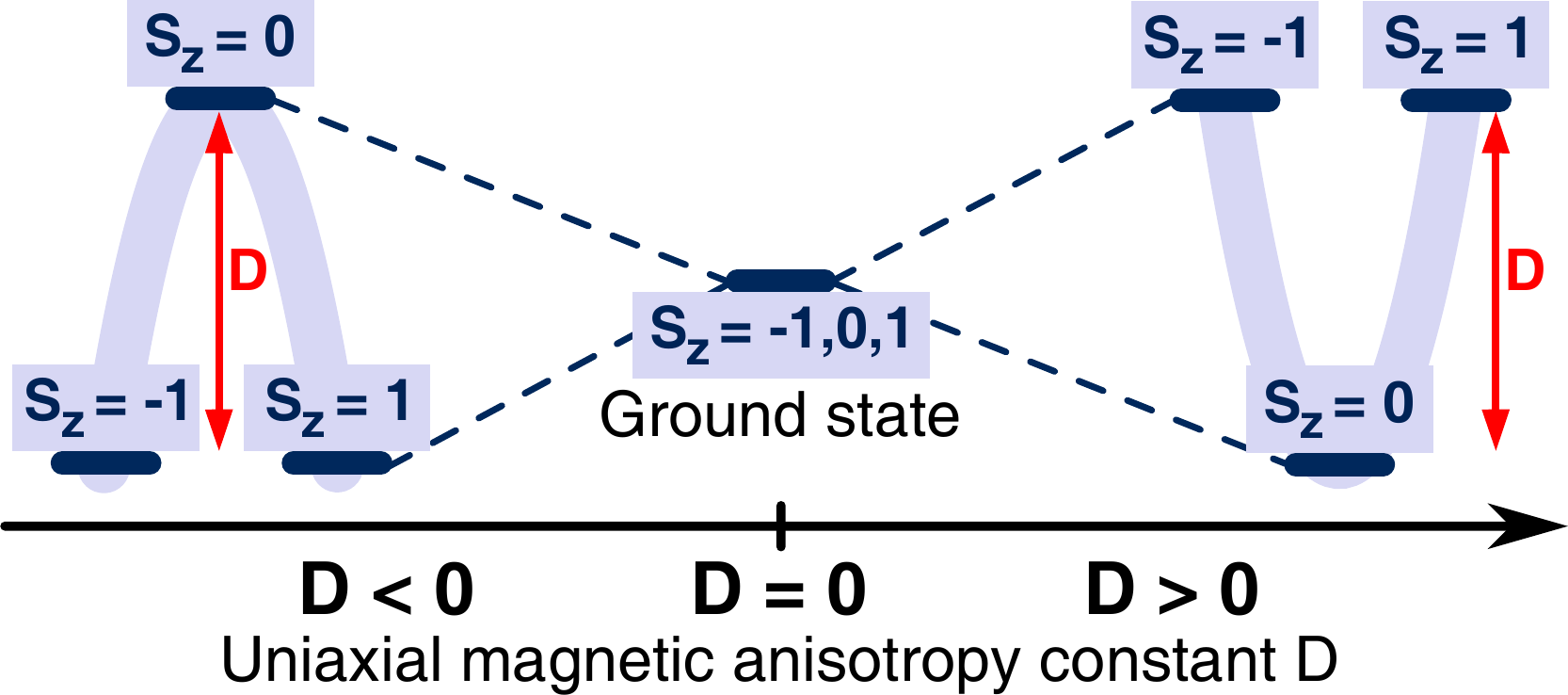}
  \caption{\label{Fig:2} (Color online)
  Sketch showing how the uniaxial magnetic anisotropy lifts  partially the degeneracy between
  the components of the triplet $S=1$.
  Note that in reality the energy of the $S_z=0$ component
  is independent of magnetic anisotropy $D$.}
\end{figure}

\subsection{Isolated quantum dot}

Before presenting and discussing numerical results on the spectral
functions, it is advisable to have a closer look at the energy
spectrum of an isolated quantum dot. For the sake of clarity of
the following discussion, let us assume that there is no external
magnetic field, $B_z=0$, so that the system's behavior is
entirely determined by both the sign and magnitude of the uniaxial
anisotropy constant $D$, as shown schematically in
Fig.~\ref{Fig:2}.

In order to observe the underscreened Kondo effect, the quantum
dot needs to be occupied by two  electrons which are
\emph{ferromagnetically} exchange-coupled. The ground state of the
dot is then a triplet with the components,
    \begin{equation}\label{Eq:8}
    \left\{
        \begin{aligned}
        &|T_+\rangle\equiv|S_z=+1\rangle=|\!\!\uparrow\rangle_1|\!\!\uparrow\rangle_2,
        \\
        &|T_0\rangle\equiv|S_z=0\rangle=\frac{1}{\sqrt{2}}\Big[|\!\!\uparrow\rangle_1|\!\!\downarrow\rangle_2+|\!\!\downarrow\rangle_1|\!\!\uparrow\rangle_2\Big],
        \\
        &|T_-\rangle\equiv|S_z=-1\rangle=|\!\!\downarrow\rangle_1|\!\!\downarrow\rangle_2,
        \end{aligned}
    \right.
    \end{equation}
where $|\chi\rangle_j$ denotes the local state of the $j$th level,
with $\chi=0,\downarrow,\uparrow,d$ corresponding to
zero, spin-down, spin-up and two electrons occupying the level,
respectively. As long as an external magnetic field and the
magnetic anisotropy are absent, the three triplet states remain
degenerate and
$\varepsilon_{T_+}=\varepsilon_{T_0}=\varepsilon_{T_-}=2\varepsilon+U+2J$,
see Fig.~\ref{Fig:2}. Moreover, the triplet remains the ground
state provided the condition,
$\delta/2-2U+5J/4<\varepsilon<-\delta/2-U-5J/4$, is
satisfied.~\cite{Weymann_Phys.Rev.B81/2010} However, the magnetic anisotropy $D$ lifts this degeneracy and the triplet
becomes partially split,
    \begin{equation}\label{Eq:9}
    \left\{
        \begin{aligned}
        &\varepsilon_{T_+}\equiv \e_{T_1}=2\varepsilon+U+2J+D,
        \\
        &\varepsilon_{T_0} = 2\varepsilon+U+2J,
        \\
        &\varepsilon_{T_-}\equiv \e_{T_1} = 2\varepsilon+U+2J+D.
        \end{aligned}
    \right.
    \end{equation}
As one can see, the energy of states $|T_+\rangle$ and  $|T_-\rangle$ depends on $D$, while the energy
of $|T_0\rangle$ is independent of $D$. Consequently, when $D<0$,
the ground state is two-fold degenerate and corresponds to the
states $|T_+\rangle$ and  $|T_-\rangle$, while for $D>0$, the
ground state corresponds to $|T_0\rangle$, see Fig.~\ref{Fig:2}.
This will have a large impact on the Kondo effect, as discussed later on. Note that the
presence of magnetic field additionally splits the states $|T_+\rangle$ and
$|T_-\rangle$.

\subsection{Effective exchange field and anisotropy\label{Sec:Beff_and _Deff}}

\begin{figure}[t]
 \includegraphics[width=0.95\columnwidth]{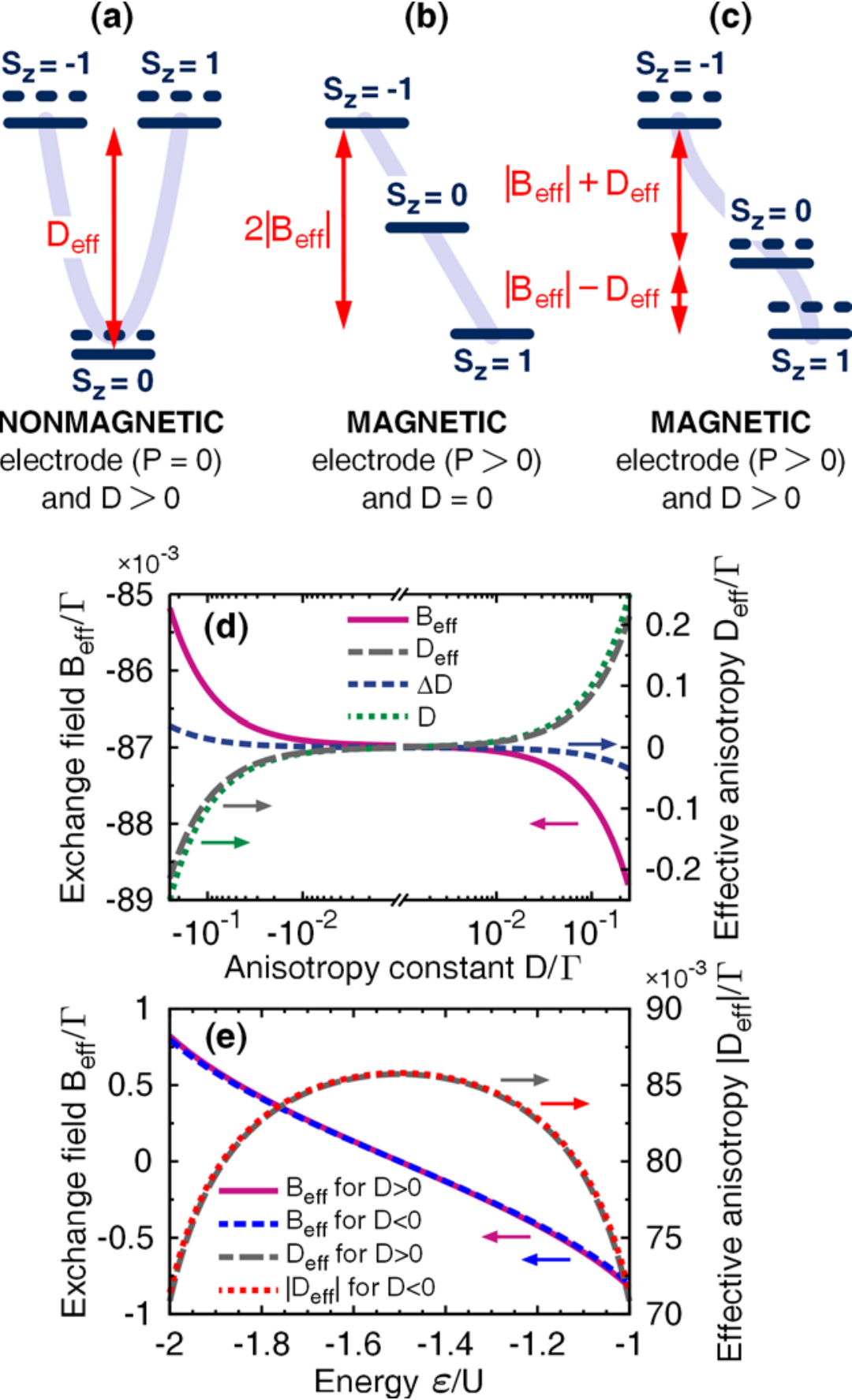}
  \caption{\label{Fig:3} (Color online)
(a)--(c) Schematic representing level renormalization due to the
effective anisotropy constant $D_{\rm eff}$ and effective exchange field $B_{\rm eff}$ for
the quantum dot with $S=1$. Dashed lines in (a) and (c) correspond
to energy levels without the renormalization of the uniaxial
magnetic anisotropy constant, $D_{\rm eff} = D$, see Eq.~(\ref{Eq:Deff}).
It is assumed that $B_{\rm eff}<0$,
so that the ground state (in (b) and (c)) is $\ket{T_+}$. Bottom
panel: the dependence of $B_{\rm eff}$ and $D_{\rm eff}$ on the
anisotropy constant $D$ for $\varepsilon/U=-1.4$ (d) and on the
level position $\varepsilon$ in the Coulomb blockade regime for
$|D|/\Gamma=0.1$ (e). Note that in (e) for $D<0$ one also gets
$D_{\rm eff}<0$, but for practical reasons  $|D_{\rm eff}|$ is
plotted.  For comparison, in (d) we also show $\Delta D$ and $D$.
The other parameters are: $\varepsilon=-17.5\Gamma$,
$\delta=2.5\Gamma$, $U=12.5\Gamma$, $J=-2.5\Gamma$, and $P=0.5$.}
\end{figure}

The next question that straightforwardly arises is what happens
when the $S=1$ quantum dot becomes attached to a reservoir of
electrons. If the temperature is lower than the Kondo temperature, $T<T_K$, the
conduction electrons can then screen only a half of the dot's
spin, whereas the residual spin-one-half is left unscreened. At
zero temperature, the system behaves then as a singular Fermi
liquid, i.e. the Fermi liquid with a decoupled $S=\frac{1}{2}$
object.~\cite{Koller_PhysRevB.72.045117,Mehta_Phys.Rev.B72/2005} In addition, it turns out
that multiple spin-flip processes responsible for the Kondo
resonance lead to renormalization of the quantum dot parameters.
One can in principle distinguish two different effects associated
with such a renormalization. First, the spin degeneracy of the dot
is lifted by an effective tunnel-induced exchange field
$B_\textrm{eff}$.~\cite{Martinek_Phys.Rev.Lett.91/2003_127203,Martinek_Phys.Rev.Lett.91/2003_247202,
Choi_Phys.Rev.Lett.92/2004,Braun_Phys.Rev.B70/2004,Martinek_Phys.Rev.B72/2005,Koenig_Lect.NotesPhys.658/2005,
Sindel_Phys.Rev.B76/2007,Weymann_PhysRevB.83.113306,Zitko_Phys.Rev.Lett.108/2012}
It was shown that the exchange field can be tuned by a gate
voltage~\cite{Hauptmann_Nat.Phys.4/2008}
and can be compensated by applying an external magnetic
field.~\cite{Pasupathy_Science306/2004,Weymann_Phys.Rev.B81/2010,
Hauptmann_Nat.Phys.4/2008,Gaass_Phys.Rev.Lett.107/2011}
The second effect, on the other hand, is related to the renormalization of the
anisotropy constant, $\Delta D$, which results in
an effective anisotropy $D_{\rm eff}$, $D_{\rm eff} = D+\Delta D$,
as shown schematically in Figs.~\ref{Fig:3}(a)-(c). This renormalization is independent of
the spin polarization $P$ of the lead, while it depends weakly on
the gate voltage and cannot be compensated by external magnetic field.

The renormalized energies $\widetilde{\varepsilon}_{T_i}$ ($i=0,\pm$) of the triplet state can be found
from the second-order perturbation theory in the tunneling Hamiltonian as
$\widetilde{\varepsilon}_{T_i} = \varepsilon_{T_i} + \delta \varepsilon_{T_i}$ ($i=0,\pm$),
where $\delta \varepsilon_{T_i}$ is the second-order correction of the respective triplet component.
The renormalized triplet energies may be written in the following way
    \begin{equation}\label{Eq:10}
    \left\{
        \begin{aligned}
        &\widetilde\varepsilon_{T_+}=\varepsilon_{T_0} + \delta\varepsilon_{T_0} + D_{\rm eff} + B_{\rm eff} + B_z,
        \\
        &\widetilde\varepsilon_{T_0}=\varepsilon_{T_0} + \delta\varepsilon_{T_0},
        \\
        &\widetilde\varepsilon_{T_-}=\varepsilon_{T_0} + \delta\varepsilon_{T_0} + D_{\rm eff} - B_{\rm eff} - B_z,
        \end{aligned}
    \right.
    \end{equation}
where the \emph{effective exchange field} $B_\textrm{eff}$ induced
by a ferromagnetic contact is given by~\cite{Martinek_Phys.Rev.B72/2005}
    \begin{equation}\label{Eq:Beff}
    B_\textrm{eff}=-\frac{\Gamma P}{\pi}\!
    \sum_{j=\pm}\!\int^\prime\!\!\! \textrm{d}\omega\,
    \Bigg\{
        \frac{1-f(\omega)}{\omega-E_{T_1,1j}}+
        \frac{f(\omega)}{\omega+E_{T_1,3j}}
    \Bigg\},
    \end{equation}
and the \emph{effective anisotropy} $D_\textrm{eff}$ can be expressed as
    \begin{equation}\label{Eq:Deff}
    D_\textrm{eff}=D+\Delta D
    \end{equation}
with the renormalization of the anisotropy constant $\Delta D$ of the form
    \begin{align}\label{Eq:Delta_D}
    \hspace*{-1pt}
    \Delta D=\frac{\Gamma}{\pi}\!\sum_{j=\pm}\!\int^\prime\!\!\!\textrm{d}\omega\,
    \Bigg\{
    &\frac{1-f(\omega)}{\omega-E_{T_0,1j}}-\frac{1-f(\omega)}{\omega-E_{T_1,1j}}
            \nonumber\\
    -\Bigg[&\frac{f(\omega)}{\omega+ E_{T_0,3j}}-\frac{f(\omega)}{\omega+ E_{T_1,3j}}
    \Bigg]
    \Bigg\}.
    \end{align}
The prime superscript in the above equations symbolizes Cauchy's principal
value integrals, and $f(\omega)$ stands for the Fermi-Dirac distribution function of the contact.
We note that terms involving  $1-f(\omega)$ represent here electron-like charge fluctuations,
due to which the molecule loses one electron, while terms with $f(\omega)$ refer to the hole-like processes,
when the charge of the molecule is increased by one electron.
Furthermore, $E_{\alpha, \beta} = \e_\alpha - \e_\beta$ is the energy
difference between the corresponding states. The respective
energies of singly occupied states are, $\e_{1\pm} = \e \pm
\delta/2 +3J/4 +D/4$, while the energies of states with three
electrons are given by, $\e_{3\pm} = 3 \e +3U \pm \delta/2 +3J/4
+D/4$.
In Eq.~(\ref{Eq:10}) $\delta\varepsilon_{T_0}$ ($\delta\varepsilon_{T_0}<0$) denotes
the second-order energy correction of the triplet component
$\ket{T_0}$ (uniform shift of the whole triplet). The explicit
form of $\delta\varepsilon_{T_0}$ is not relevant for the present
discussion, since it is the difference between the above energies of triplet components
that determines the occurrence and features of the Kondo effect.
In the \emph{low temperature} regime,
which except Sec.~\ref{Sec:Temperature_dependence} is in the main scope of the work,
Eqs.~(\ref{Eq:Beff}) and~(\ref{Eq:Delta_D}) simplify significantly,
    \begin{equation}\label{Eq:11}
      B_{\rm eff} = \frac{P\Gamma}{\pi}  \ln\left| \frac{ E_{T_1,1-} } { E_{T_1,3-} }\cdot
      \frac{E_{T_1,1+}} {E_{T_1,3+}} \right| ,
    \end{equation}
    \begin{equation}\label{Eq:12}
    \Delta D = \frac{\Gamma}{\pi} \ln \left|
    \frac{ E_{T_1,1-} } { E_{T_0,1-} }\cdot
    \frac{ E_{T_1,1+} } { E_{T_0,1+} }\cdot
    \frac{ E_{T_1,3-} } { E_{T_0,3-} }\cdot
    \frac{ E_{T_1,3+} } { E_{T_0,3+} }
    \right|.
    \end{equation}
We also note that the energies of triplet, Eq.~(\ref{Eq:10}), explicitly include
the external magnetic field $B_z$. This will be relevant for the discussion
of system transport properties in the presence of magnetic field,
that will be presented in the next section.

From Eqs.~(\ref{Eq:Beff}) and (\ref{Eq:11}) follows that the exchange
field is an intrinsic effect resulting from the spin-dependence of
tunneling processes, and vanishes for $P\to 0$. Moreover, $B_{\rm
eff}$ displays monotonic dependence on the level position
$\varepsilon$: with $B_{\rm eff}=0$ at the particle-hole symmetry
point of the model, i.e. for $\e = -3U/2$, and $B_{\rm eff}
\lessgtr 0$ for $\e \gtrless -3U/2$. This is shown in
Fig.~\ref{Fig:3}(e). In addition, $B_{\rm eff}$ also depends on
the anisotropy constant $D$, see Fig.~\ref{Fig:3}(d). This
dependence, however, is much weaker than the dependence on
$\varepsilon$.

On the other hand, since $\e_{T_1} = \e_{T_0}+D$, one can immediately conclude from
Eqs.~(\ref{Eq:Delta_D}) and (\ref{Eq:12}) that $\Delta D \to 0$ as $D \to 0$.
It is interesting to note that tunneling of electrons leads to
suppression of the magnetic anisotropy, $\Delta D\lessgtr 0$ for
$D\gtrless0$, in the whole Coulomb blockade regime. Furthermore,
unlike the effective exchange field, $\Delta D$ is
an even function of the level position $\varepsilon$ with the
extremum (maximum for $D>0$ and minimum for $D<0$) in the
particle-hole symmetry point, where
    \begin{equation}\label{Eq:13}
    \Delta D_{\e = -\frac{3U}{2}} = \frac{\Gamma}{\pi} \ln \left|
    \frac{ (2U-5J-3D)^2 - 4\delta^2 } { (2U-5J+D)^2 - 4\delta^2 }
    \right|.
    \end{equation}
The dependence of the effective anisotropy $D_{\rm eff}$ on the
magnetic anisotropy constant as well as on the level position is
shown in Fig.~\ref{Fig:3}(d,e). One can note that $D_{\rm eff}$
depends strongly on $D$ and only weakly on the level position,
which is just opposite to the behavior of the exchange field
$B_{\rm eff}$. Moreover, while $B_{\rm eff}$ is due to the
spin-dependence of tunneling processes and vanishes for
nonmagnetic leads, $D_{\rm eff}$ does not depend on the spin
polarization and is finite also when $P=0$ -- as long as $D\neq
0$.

The above discussion suggests that transport properties should be
mainly determined by the interplay of the effective anisotropy $D_{\rm eff}$,
contact-induced exchange field $B_{\rm eff}$ and the Kondo temperature $T_K$.
Additionally, the behavior of the total spectral function also
significantly depends on the tunnel-coupling strength $\Gamma$ (to
observe the Kondo physics the coupling should be sufficiently
large). Experimental results show that the Kondo phenomena  in
quantum dots can be observed when $\Gamma$ is of the order of a
few tenths of meV for temperatures of the order of mK.
~\cite{Goldhaber-Gordon_Nature391/1998,Simmel_Phys.Rev.Lett.83/1999,
Sasaki_Nature405/2000,Kogan_Phys.Rev.B67/2003} Accordingly, in
numerical calculations we assume $\Gamma=0.5$ meV and use $\Gamma$
as the relevant energy scale. For the quantum dot we assume the
parameters that are comparable to those observed in experiments,
$\varepsilon=-17.5\Gamma$ ($\varepsilon=-8.75$ meV),
$\delta=2.5\Gamma$ ($\delta=1.25$ meV), $U=12.5\Gamma$ ($U=6.25$
meV), and $J=-2.5\Gamma$ ($J = -1.25$ meV), if not stated
otherwise. Note that we assumed $\varepsilon/U
> -3U/2$, so that if the lead is ferromagnetic, the ground state is
$\ket{T_+}$ due to $B_{\rm eff}<0$, see also Fig.~\ref{Fig:3}(c).

\section{Numerical results and discussion}

In the following we present and discuss numerical results on the
dot's spectral density as a function of the anisotropy constant $D$,
spin polarization of the lead $P$, and external magnetic field $B_z$. The
main focus, however, will be on the effects arising from the
magnetic anisotropy and effective exchange field. Generally, the
magnetic anisotropy in systems under consideration can take fairly
large values, and can range
approximately from $|D|\lesssim0.05$ meV for single-molecule
magnets, ~\cite{Gatteschi_book,Heersche_Phys.Rev.Lett.96/2006,
Zyazin_NanoLett.10/2010,Mannini_Nature468/2010} up to a few meV
for magnetic adatoms like Mn, Fe, Co,
~\cite{Hirjibehedin_Science317/2007,Otte_NaturePhys.4/2008,Brune_Surf.Sci.603/2009}
or some magnetic molecules.~\cite{Parks_Science328/2010}
Furthermore, Park \emph{et al.}~\cite{Parks_Science328/2010} have
shown that mechanical strain in a certain type of Co complexes
allows for a fully controllable and continuous change of the
magnetic anisotropy of a molecule. This effect occurs since the
stretching or squeezing of a molecule leads to modification of the
crystal field exerted on the central Co ion. For the above
reasons, the following results will be presented for a wide range
of both positive and negative uniaxial anisotropy constant $D$.

\subsection{Influence of uniaxial magnetic anisotropy}

\begin{figure}[t]
  \includegraphics[width=0.99\columnwidth]{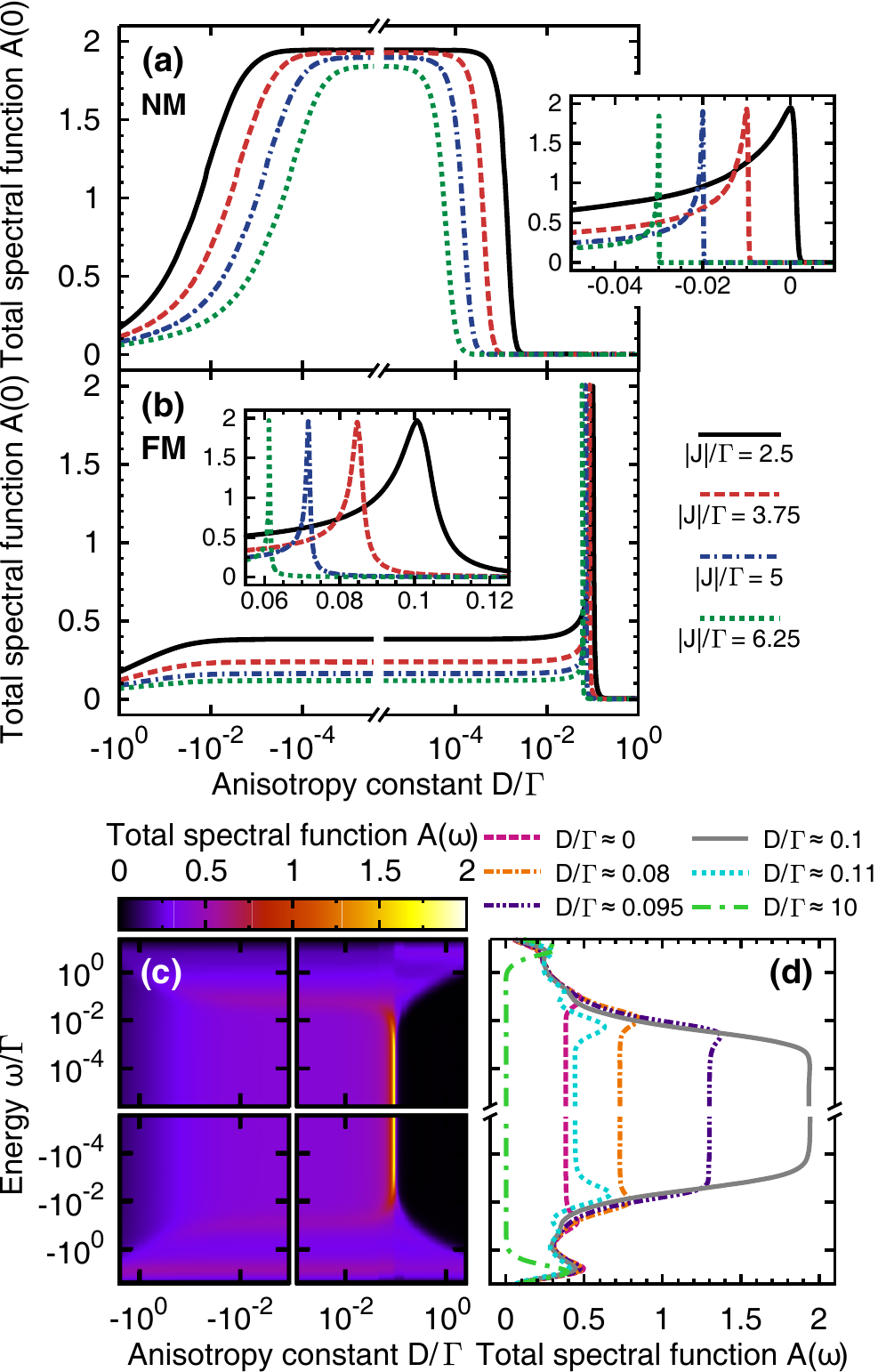}
  \caption{\label{Fig:4} (Color online) Total \emph{normalized} spectral function
  $A(\omega)$ of the dot for $\omega\rightarrow0$, $A(0)$, as a function of
  the uniaxial magnetic anisotropy constant $D$ for several values of the
  interlevel exchange interaction $J$ in the presence of
  (a) \emph{nonmagnetic} (NM) and (b) \emph{ferromagnetic} (FM) electrode.
  The insets in (a) and (b) display $A(0)$ as a function of $D/\Gamma$
  plotted in a linear scale in the case of nonmagnetic (a) and ferromagnetic (b) electrode
  for different values of $J$. Note that in the inset in (a) the Kondo peaks have been shifted by $-0.01$ for better visibility.
  (c) Variation of the spectral function with the energy $\omega$ and uniaxial
  magnetic anisotropy constant $D$ for $J/\Gamma=-2.5$.
  (d) Cross-sections of the plot in (c) for selected values
  of the uniaxial anisotropy constant $D>0$.
  The other parameters are: $\varepsilon=-17.5\Gamma$,
  $\delta=2.5\Gamma$, $U=12.5\Gamma$, and $P=0.5$ (except (a) where $P=0$).}
\end{figure}

\subsubsection{Nonmagnetic lead}

To begin with, let us first discuss briefly how the spectral
function of the system depends on the magnetic anisotropy in the
case of a \emph{nonmagnetic} electrode.~\cite{Cornaglia_Europhys.Lett.93/2011} Figure~\ref{Fig:4}(a)
shows the dependence of $A(0)$ on $D$ for several values
of the exchange interaction $J$. It can be seen that
$A(0)\approx 2$ for $|D| < T_K$, where $T_K$ is the Kondo temperature,
with $A(0)$ decreasing below its unitary value once $|D| \gtrsim T_K$.
Moreover, the resonance dies away more abruptly for $D>0$, where the
spectral function is practically equal to zero above some
threshold value of the anisotropy constant. Accordingly, one
should expect there a vanishingly small linear conductance of the
system. Indeed, such a behavior has been observed by Parks
\emph{et al.},~\cite{Parks_Science328/2010} who reported splitting
of the Kondo peak due to stretching the molecule. The asymmetry
between the decrease of $A(0)$ for positive and negative $D$ is
associated with different ground states of the quantum dot, see
Fig.~\ref{Fig:2}. For  $D>0$ and $D_{\rm eff} \gtrsim
T_K$, the ground state is $\ket{T_0}$ and no spin-flip processes
are possible, consequently $A(0)$ becomes abruptly suppressed. On the
other hand, for  $D<0$ and $|D_{\rm eff}|\gtrsim T_K$, the
ground state is two-fold degenerate, with equally occupied states
$\ket{T_+}$ and $\ket{T_-}$, see Eq.~(\ref{Eq:10}).
Due to the spin selection rules for
tunneling processes, the second-order spin-flip cotunneling is
then suppressed and the Kondo resonance becomes suppressed as well
[$A(0)$ starts decreasing]. However, there are fourth-order
tunneling processes that are still possible and therefore $A(0)$
decreases rather slowly with increasing $|D|$, opposite to the
case of positive $D$.

The features discussed above depend on the Kondo temperature
$T_K$. Since $T_K$ is a function of the energy difference between
the ground state and single and three-particle virtual states, the
Kondo temperature can be tuned by changing the exchange
interaction $J$. When increasing $|J|$, one effectively increases
the energy differences and $T_K$ becomes decreased.
Accordingly, the suppression of the Kondo effect occurs
for smaller values of $D$. This can be clearly seen in
Fig.~\ref{Fig:4}(a). Note that the width of the maximum in $A(0)$
as a function of  $D$ is roughly equal to $2T_K$.

\subsubsection{Ferromagnetic lead}

The situation becomes much more interesting when the
nonmagnetic reservoir is replaced by the ferromagnetic one.
The dependence of the spectral function $A(0)$ on $D$ is presented in Fig.~\ref{Fig:4}(b).
First, for small
values of $|D|$, and thus also $|D_{\rm eff}|$, the height of the Kondo resonance is significantly
reduced as compared to the case of a nonmagnetic electrode, which
is due to the presence of exchange field $B_{\rm eff}$. Second, as $D>0$
increases, one observes the revival of the Kondo effect at some
resonant value of the magnetic anisotropy constant,
$D=D_\textrm{res}$. However, further increase in $D$ results in the
drop of the spectral function to zero, so that the behavior of the
system for large and positive $D$ resembles that of a dot coupled
to a nonmagnetic electrode. In addition, the dependence on the
exchange coupling is also qualitatively similar to that in the
nonmagnetic case. When increasing $|J|$,
$T_K$ is reduced and the width of the Kondo resonance as a function of $D$
becomes decreased as well, see the inset in Fig.~\ref{Fig:4}(b).
In fact, the width of the Kondo resonance for a given value of $J$
is  of the same order in the case of nonmagnetic and ferromagnetic leads,
compare the insets in Figs.~\ref{Fig:4}(a,b).
In addition, the value of $D_{\rm res}$ decreases with increasing
$J$, which is due to the corresponding dependence of $B_{\rm eff}$
and $D_{\rm eff}$ on the exchange coupling $J$.

Since the spectral function can be substantially modified upon
altering the anisotropy constant $D$, the following discussion
will be focused on the interplay of $D_{\rm eff}$ and $B_{\rm
eff}$, that governs the transport behavior in the Kondo regime. In
the remaining part of the paper we will present and discuss
numerical results for a fixed value of the exchange coupling,
$J/\Gamma=-2.5$ [corresponding to the bold lines in
Figs.~\ref{Fig:4}(a,b)].

\begin{figure}[t]
  \includegraphics[width=0.75\columnwidth]{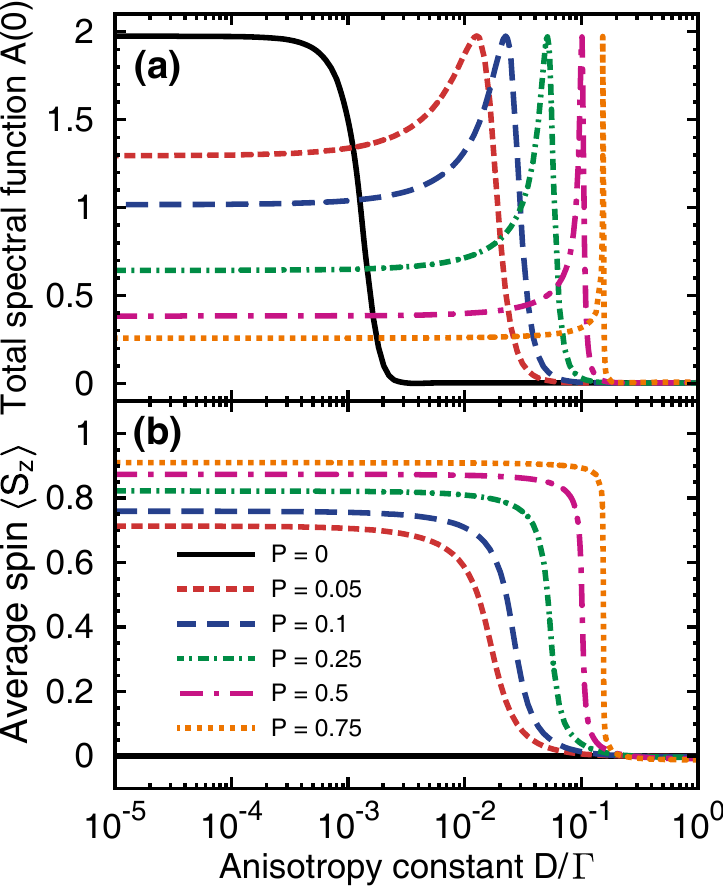}
  \caption{\label{Fig:5} (Color online) (a)
  Total \emph{normalized} spectral function $A(\omega)$ of the dot for $\omega\rightarrow0$
  and (b) the corresponding expectation value of the dot's spin $z$th component
  as functions of the uniaxial magnetic anisotropy constant $D$.
  Except $J/\Gamma=-2.5$, remaining parameters are the same as in Fig.~\ref{Fig:4}.
}
\end{figure}

As follows from Eq.~(\ref{Eq:11}), the strength of $B_\textrm{eff}$ depends
on the spin polarization $P$ of the reservoir, $B_\textrm{eff} \sim P \Gamma$.
Moreover, through the energy differences between respective states,
$B_\textrm{eff}$ is a function of all parameters of the model,
including the magnetic anisotropy constant $D$.
First of all, unlike $\Delta D$ [see Eq.~(\ref{Eq:12})], $B_\textrm{eff}$ is finite for $D\rightarrow 0$.
More specifically, for parameters used in Figs.~\ref{Fig:3}(d) and
\ref{Fig:4}(c,d) it approaches a constant value of
$B_\textrm{eff}/\Gamma\approx-0.087$.
Since the exchange field lowers the energy of the highest-weight triplet component $\ket{T_+}$,
see schema (b) in Fig.~\ref{Fig:3},
one can observe almost full spin polarization of the dot,
$\expect{ S_z} \to 1$, see Fig.~\ref{Fig:5}(b) for $D\to 0$.
Note, however, that the dot's spin can be flipped to $\expect{S_z}\to -1$
for $\e<-3U/2$, where $B_{\rm eff}>0$ and $\ket{T_-}$ becomes the
ground state of the system. This can be achieved for instance by
applying a gate voltage. In addition, $B_{\rm eff}$ also depends
on the anisotropy constant and it can either increase or
decrease depending on the sign of $D$, see the solid line in
Fig.~\ref{Fig:3}(d). For the parameters used in calculations, the
modification of $B_{\rm eff}$ for $|D| = \Gamma/10$ is
however rather small ($\sim1\%$). The variation of the exchange field
as a function of $D$ is thus of rather minor
significance for $\Gamma\gg|D|$, but nevertheless the interplay of
$D$ ($D_{\rm eff}$) and $B_\textrm{eff}$ turns out to be crucial
for the occurrence of the Kondo effect.

Let us focus first on the case of $D<0$, where no restoration of
the Kondo resonance takes place, see the left side of
Fig.~\ref{Fig:4}(b). The ground state for $P=0$  would be doubly
degenerate. Because of the exchange field, this degeneracy,
however, is lifted and the ground state is $|T_+\rangle$. The
state $|T_+\rangle$ remains the ground state in the whole range of
$D<0$ considered in this paper. In consequence, there is no Kondo
effect for $D<0$.

The situation, however, is  much more complex in the case of
$D>0$, where the restoration of the Kondo resonance appears, see
the right side of Fig.~\ref{Fig:4}(b). For small values of $D$,
where the dominant energy scale due to renormalization processes
is set by the exchange field $B_\textrm{eff}$, the ground state is
$\ket{T_+}$ and the situation is similar to that for $D<0$.
Nonetheless, as the magnetic anisotropy grows, the condition
$B_\textrm{eff} + D_\textrm{eff}=0$ becomes satisfied at some
point (note that for the assumed parameters $B_{\rm eff}<0$ while
$D_{\rm eff}>0$) and the state $|T_+\rangle$ gets degenerate with
the state $|T_0\rangle$. The difference between the spin $z$th
components of the states $\ket{T_+}$ and $\ket{T_0}$ is 1, so the
second-order spin-flip cotunneling processes are possible and the
Kondo effect can be restored. $A(0)$ reaches then its maximal value.

\begin{figure}[t]
  \includegraphics[width=0.99\columnwidth]{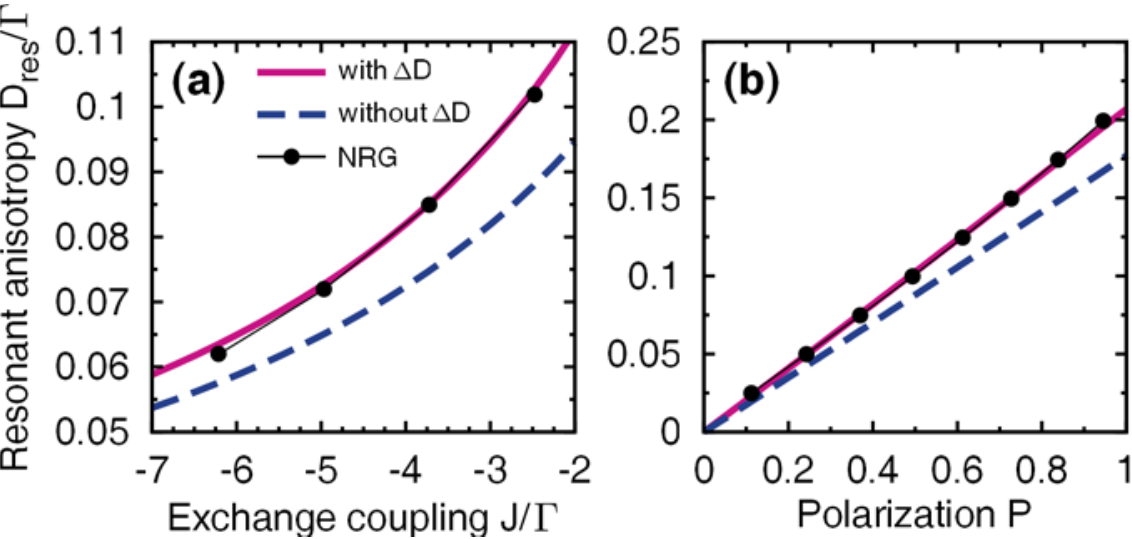}
  \caption{\label{Fig:6} (Color online)
  (a) Dependence of the magnetic anisotropy constant $D_\textrm{res}$
  at which the Kondo resonance is restored on the interlevel exchange
  coupling $J$ for $P=0.5$ and
  (b) on the spin polarization $P$ for $J/\Gamma=-2.5$.
  Other parameters as in Fig.~\ref{Fig:4}.}
\end{figure}

\begin{figure*}
  \includegraphics[width=1.85\columnwidth]{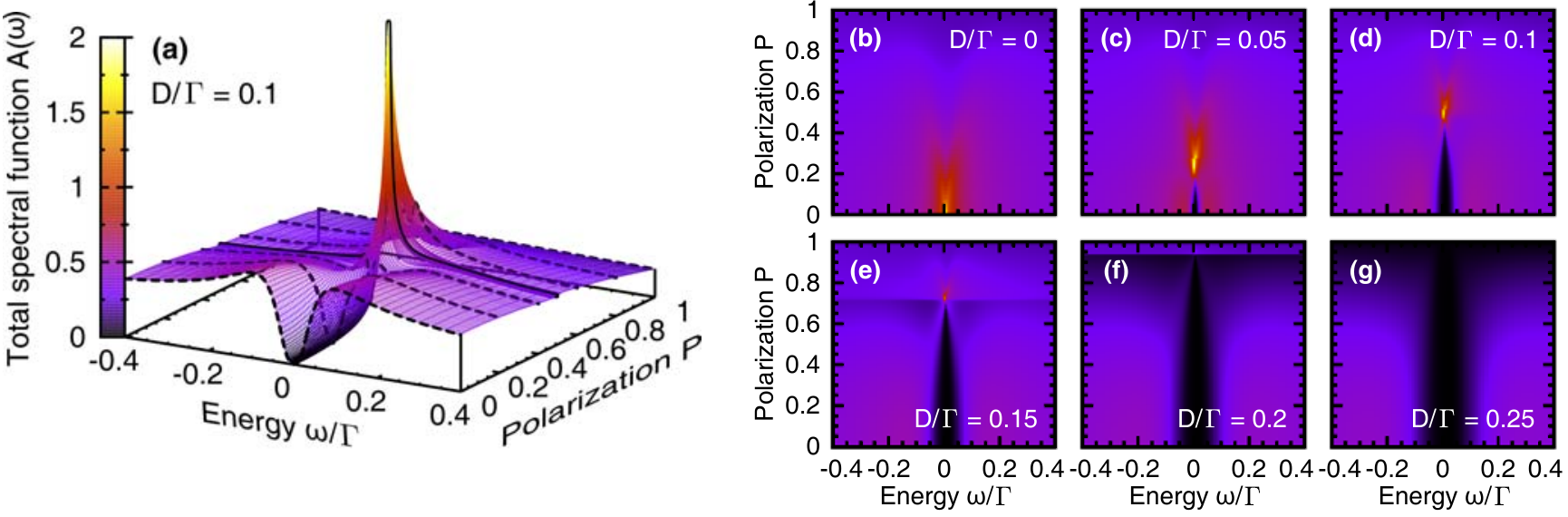}
  \caption{\label{Fig:7} (Color online)
  Total \emph{normalized} spectral function $A(\omega)$ of the dot
  shown as a function of energy $\omega$ and the spin polarization
  parameter $P$ for indicated values of the uniaxial magnetic
  anisotropy constant $D$.
  Other parameters are the same as in Fig.~\ref{Fig:5}.
  }
\end{figure*}

The above described behavior can be also observed in the full
energy dependence of the spectral function $A(\omega)$, see
Figs.~\ref{Fig:4}(c,d). The Kondo resonance is restored when $D =
D_{\rm res}\approx \Gamma/10$ and is immediately suppressed once
$D>D_{\rm res}$. On the other hand, the suppression is less
effective on the left side of the restored Kondo resonance, as
already discussed above. In addition, the energy dependence of the
spectral function in Fig.~\ref{Fig:4}(d) reveals  small side peaks
for $D>D_{\rm res}$, which are reminiscent of the Kondo effect and
occur for energies $\omega \approx \pm |D_{\rm eff}+B_{\rm eff}|$
corresponding to restored degeneracy of the states $\ket{T_+}$ and
$\ket{T_0}$. Apart from this, at large energies, $\omega \approx
\pm U$, there are typical Hubbard resonance peaks.

Since the occurrence of the Kondo resonance depends on
the ratio of $B_{\rm eff}$ and $D_{\rm eff}$, it is interesting to study
variation of $A(0)$ with $D$ for different values of lead's spin polarization.
This is shown in Fig.~\ref{Fig:5}(a).
When $P=0$, $A(0)$ shows a maximum for such $D$ that
the condition $|D_{\rm eff}|\lesssim T_K$ is fulfilled. If the spin
polarization is finite, the maximum is shifted towards larger
values of anisotropy and occurs precisely when the states $\ket{T_+}$ and
$\ket{T_0}$ become degenerate, i.e. for $D=D_{\rm res}$. Note,
that the width of the peak in $A(0)$ as a function of $D$ is of
the same order for all values of spin polarization $P$.
Since the dot is coupled to one electron reservoir, only
half of the dot's spin can be screened by the conduction electrons
for $D=D_{\rm res}$. As a result, in the underscreened Kondo
regime the expectation value of the dot's spin should reach
$\langle S_z\rangle \approx 1/2$, since the unscreened residual
spin-1/2 is polarized due to the presence of exchange field.
This can be seen in Fig.~\ref{Fig:5}(b), which shows the
dependence of $\expect{S_z}$ on $D$ for several values of $P$.
For $D<D_{\rm res}$, the ground state of the dot is $\ket{T_+}$ and
$\expect{S_z}$ is close to one, while for $D>D_{\rm res}$, the
ground state is $\ket{T_0}$ and $\expect{S_z}=0$. On the other
hand, for $D\approx D_{\rm res}$, one finds $\expect{S_z}=1/2$,
see Fig.~\ref{Fig:5}(b).
However, closer analysis of $\expect{S_z}$
shows that it actually fails in attaining its
maximum value for $D<D_\textrm{res}$. This is related with the
fact that the ratio $U/\Gamma$ is relatively large for the assumed
parameters and there is nonzero occupation probability of other
spin components of the triplet.~\cite{Weymann_Phys.Rev.B81/2010}
However, when increasing the spin polarization $P$ of the lead,
the splitting of the levels grows due to the exchange field,
$B_{\rm eff} \sim P\Gamma$, and the occupation of the triplet component
$\ket{T_+}$ is raised. In consequence, one finds that
$\expect{S_z} \to 1$, if $P\to 1$.

Knowing the analytical condition, $B_\textrm{eff} + D_\textrm{eff}=0$,
for the occurrence of the Kondo resonance for
$D>0$, it is instructive to analyze the role of magnetic anisotropy
renormalization. For this purpose, in Fig.~\ref{Fig:6} we
present the dependence of $D_\textrm{res}$ on $J$ and $P$.
The solid (dashed) line corresponds to $D_\textrm{res}$ determined from
the analytical formulas for $B_\textrm{eff}$ and $D_\textrm{eff}$
with (without) including $\Delta D$, while the dots show $D_\textrm{res}$
obtained from NRG data.
As one can see, the NRG results are in very good agreement with
analytical results when $\Delta D$ is taken into account. Thus, the
estimations based on the analytical expressions for the exchange
field and effective anisotropy, Eqs.~(\ref{Eq:11})-(\ref{Eq:12}), are quite satisfactory.
The renormalization of $D$ is thus
an important effect that needs to be included in theoretical considerations
of spin $S\gtrsim 1$ systems exhibiting magnetic anisotropy.

To demonstrate additional features of the interplay between magnetic
anisotropy and exchange field, we show in Fig.~\ref{Fig:7} the
energy and spin polarization dependence of the normalized spectral
function $A(\omega)$. The full energy dependence of the spectral
function may prove to be useful in predicting some qualitative
information concerning transport properties of the system at a
finite bias. As it was discussed earlier, the spin polarization
$P$ determines the strength of $B_\textrm{eff}$, without affecting $D_{\rm eff}$.
In consequence, all $\omega$-dependent features in
Fig.~\ref{Fig:7} should in principle stem from the changes of
$B_\textrm{eff}$ with respect to $D_\textrm{eff}$.
For $P<P_\textrm{res}$, where $P_{\rm res}$ is the value of spin
polarization at which the Kondo resonance is restored, one
observes a well-pronounced dip, which indicates that the system's
ground state is nonmagnetic, i.e. $\ket{T_0}$.
In the present picture, increasing $\omega$ turns to be
equivalent (to some extent) to the application of an external bias
voltage when the dot would be asymmetrically attached to two contacts.~\cite{Csonka2012}
For $P\rightarrow0$, the spectral function attains then its local
maximum at $\omega$ that approximately corresponds to the
degeneracy of the states $|T_0\rangle$ and $|T_+\rangle$. From this,
in turn, $D_\textrm{eff}$ can be straightforwardly
obtained, i.e. $D_\textrm{eff} \approx \omega$. As $P$ increases,
the magnitude of the exchange field $|B_\textrm{eff}|$ increases as well,
and this is accompanied by a decrease in the energy gap $B_\textrm{eff}+D_\textrm{eff}$.
This appears then as a gradual narrowing of the dip, until the two
states, i.e. $\ket{T_+}$ and $\ket{T_0}$, become degenerate, and
the Kondo resonance is restored, see Figs.~\ref{Fig:7}(a)-(e).
For larger values of $P$, the system's ground state is $|T_+\rangle$
and the Kondo resonance is suppressed again. There are however two
satellite peaks at energies $\omega\approx\pm|D_{\rm eff}+B_{\rm
eff}|$, whose position depends linearly on $P$,
and whose height diminishes as $P$ grows further. Moreover,
it turns out that for larger values of $D$, see Figs.~\ref{Fig:7}(f)-(g),
no restoration of the Kondo effect is possible.
This is because the magnitude of exchange field is too low to compensate the effective anisotropy $D_{\rm
eff}$ and the degeneracy of states cannot be restored even if $P \to 1$.

\subsection{Influence of an external magnetic field}

\begin{figure}[b]
  \includegraphics[width=0.95\columnwidth]{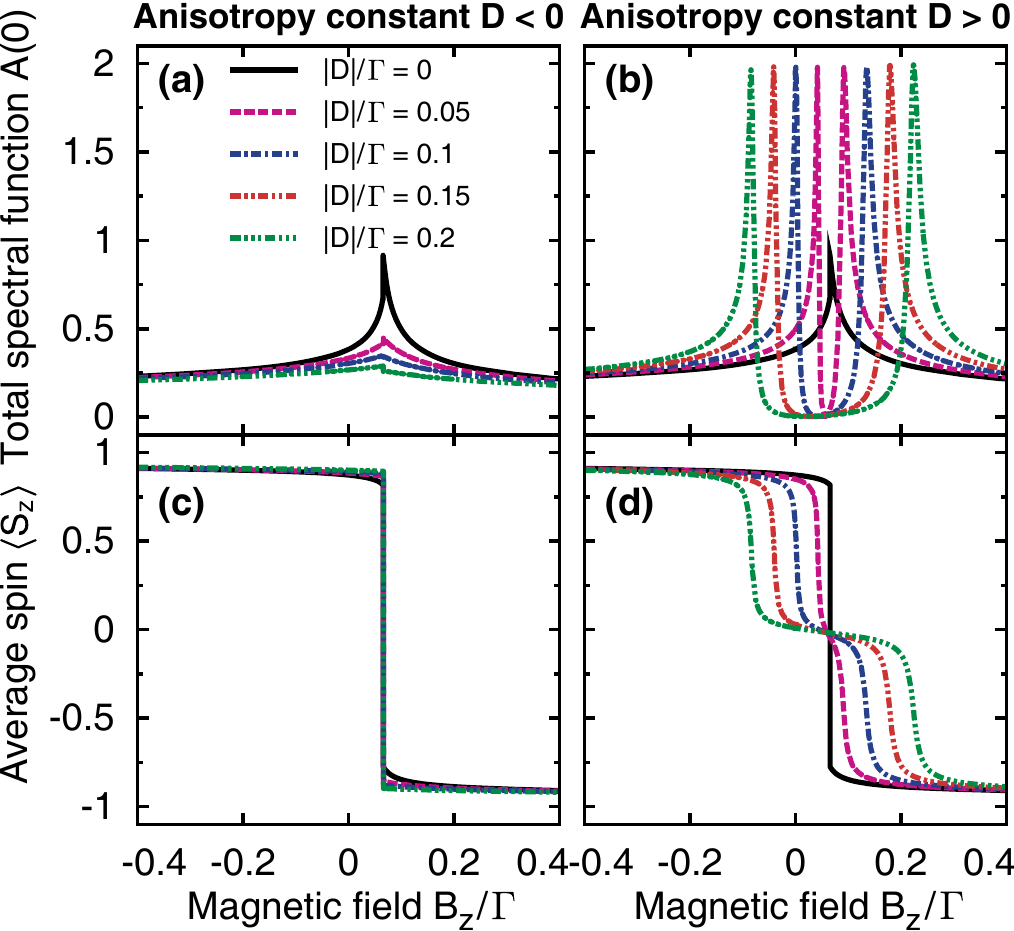}
  \caption{\label{Fig:8} (Color online)
  Dependence of the total \emph{normalized} spectral function
  $A(\omega)$ for $\omega\rightarrow0$ (a,b) and the average
  $z$th component of the spin $\langle S_z\rangle$ (c,d)
  on an external magnetic field $B_z$ oriented along the dot's easy axis.
  Different lines correspond to selected values of the uniaxial
  magnetic anisotropy constant $D$, and the left panel corresponds to $D<0$,
  while the right panel refers to $D>0$.
  Except for $P=0.5$ and $J/\Gamma=-2.5$,
  all remaining parameters are the same as in Fig.~\ref{Fig:4}.}
\end{figure}

Figure~\ref{Fig:8} shows the dependence of the
spectral function $A(0)$ and the average
value of dot's spin $\expect{S_z}$ on the magnetic field $B_z$.
It can be noticed that while the Kondo effect for $D<0$ can be restored
just for a single value of $B_z$, for $D>0$ the restoration occurs twice.
In order to understand this behavior one should bear in mind that
the external magnetic field affects the components $\ket{T_+}$
and $\ket{T_-}$ of the triplet state and thus can be used to
compensate the splitting induced by exchange field $B_{\rm eff}$
due to ferromagnetic contact,~\cite{Weymann_Phys.Rev.B81/2010} see
Eq.~(\ref{Eq:10}). In the case of vanishing $D$ and $B_z=0$, the Kondo resonance is suppressed due to the
exchange field $B_{\rm eff}$ and the system ground state is
$\ket{T_+}$. With increasing $B_z$, all the
three components of the triplet state become degenerate
once $B_{\rm eff}+B_z\approx 0$, and the system is in the underscreened
Kondo regime. Nevertheless, because only a half of the dot's spin
can be screened by conduction electrons, the remaining spin-$1/2$
can be polarized by any infinitesimal magnetic field (at zero
temperature). This leads to strong sensitivity of the ground state
on magnetic field, which hinders the full restoration of the
underscreened Kondo effect.~\cite{Weymann_Phys.Rev.B81/2010}

For $D<0$, the increase of $B_z$ can only restore the
degeneracy between the highest and lowest-weight components of the
triplet. In consequence, one observes a small peak at $B_z \approx
|B_{\rm eff}|$, see Fig.~\ref{Fig:8}(a), where the ground state changes from $\ket{T_+}$ to
$\ket{T_-}$, see Fig.~\ref{Fig:8}(c). Note that the position of
this peak does not depend on $D$. On the other
hand, for positive
anisotropy $D>0$, the full restoration of the Kondo effect
is possible, see Fig.~\ref{Fig:8}(b). Moreover, contrary to single-level quantum
dots,~\cite{Gaass_Phys.Rev.Lett.107/2011} the restoration with
increasing $B_z$ occurs twice. This can be understood by studying
the evolution of the ground state with magnetic field, see
Fig.~\ref{Fig:8}(d). For $B_z=0$ and $D>0$, the ground state is a
singlet, $\ket{T_0}$. By lowering the magnetic field, $B_z<0$, the ground state changes
to $\ket{T_+}$ for $B_z \approx - D_{\rm eff} - B_{\rm eff}$, while by increasing $B_z$, once $B_z \approx
D_{\rm eff} - B_{\rm eff}$, the ground state changes to
$\ket{T_-}$. Consequently, once $B_z \approx \pm D_{\rm eff} -
B_{\rm eff}$, the two-fold degeneracy of the ground state becomes
restored and the Kondo effect can develop. One observes then two
maxima in $A(0)$, see Fig.~\ref{Fig:8}(b). Note, however, that
different states are responsible for these two Kondo peaks. For $B_z
\approx -D_{\rm eff} - B_{\rm eff}$, it is the degeneracy between
the states $\ket{T_+}$ and $\ket{T_0}$ that results in the
formation of the Kondo effect, while for $B_z \approx D_{\rm eff}
- B_{\rm eff}$, the states $\ket{T_-}$ and $\ket{T_0}$ are
degenerate. It is also worth noting that since the Kondo
temperature is rather independent of $D$, the width of the
restored Kondo peaks is the same for all values of $D$, see
Fig.~\ref{Fig:8}(b).

From the magnetic field
dependence of $A(0)$ one can obtain the information about the
magnitude of both $B_{\rm eff}$ and $D_{\rm eff}$. Suppose the
restoration of the Kondo effect occurs for $B_z=B_{\rm res}^{(1)}$
and $B_z=B_{\rm res}^{(2)}$, then the effective anisotropy
constant can be related to a half of the distance between the two restored Kondo
resonances $D_{\rm eff} = |B_{\rm res}^{(1)}-B_{\rm
res}^{(2)}|/2$. On the other hand, the magnitude of the exchange field
can be found from the average, $B_{\rm eff} = - (B_{\rm
res}^{(1)}+B_{\rm res}^{(2)})/2$. Studying the magnetic field
dependence of the zero-energy spectral function, which would
correspond to measuring the low-temperature zero-bias conductance,
may be thus useful in obtaining information about both the
effective anisotropy and exchange field.

\subsection{\label{Sec:Temperature_dependence}Temperature dependence of linear conductance}

\begin{figure}[t]
  \includegraphics[width=0.725\columnwidth]{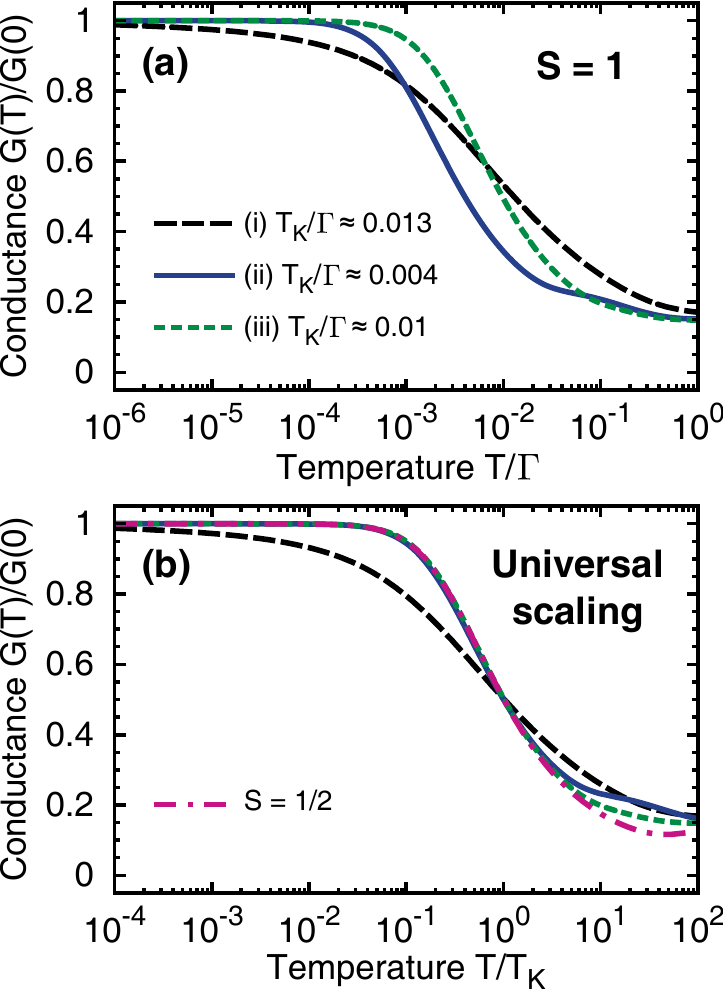}
  \caption{\label{Fig:9} (Color online)
  The temperature dependence of the normalized linear conductance $G(T)/G(0)$ (a)
  and the universal scaling curves (b).
  The relevant curves correspond to
  (i) underscreened Kondo effect, with $D=0$, $P=0$, $B_z=0$,
  (ii) restored Kondo resonance when $B_{\rm eff}+D_{\rm eff}=0$, with
  $D/\Gamma\approx0.1$, $P=0.5$, $B_z=0$, and
  (iii) restored Kondo resonance by magnetic field $B_z=D_{\rm eff}-B_{\rm eff}$,
  with $D/\Gamma\approx0.2$, $P=0.5$, $B_z/\Gamma\approx0.22$.
  The curves (ii) and (iii) display scaling typical
  for spin $S=1/2$ Kondo effect [dotted-dashed line in (b)].
  Other parameters the same as in Fig.~\ref{Fig:4} with $J/\Gamma=-2.5$.}
\end{figure}

It is very instructive to study the temperature $T$ dependence of the linear conductance, $G(T)$,
in the underscreened Kondo regime and
for parameters where the restoration of the Kondo effect appears.
Generally, the conductance in the underscreened Kondo regime can be measured by
attaching a second weakly-coupled electrode, which due to much smaller Kondo temperature,
is irrelevant for screening the dot's spin.~\cite{Roch_Phys.Rev.Lett.103/2009,Parks_Science328/2010}
The linear conductance has been calculated by means of NRG method with the full density-matrix,
and the Meir-Wingreen formula.~\cite{Meir_Phys.Rev.Lett.68/1992}
The normalized linear conductance as a function of temperature $T$ is shown in Fig.~\ref{Fig:9}(a)
for the underscreened Kondo effect (dashed line), i.e. for $P=D=B_z=0$,
and for parameters where the restoration
of the Kondo effect occurs -- first when the condition $D_{\rm eff}+B_{\rm eff} = 0$ is met
(solid line) and second when the restoration is obtained by applying magnetic field,
i.e. when $D_{\rm eff}-B_{\rm eff} - B_z= 0$ is satisfied (dotted line).
The relevant Kondo temperatures are also given in the figure. $T_K$ is defined here
as the value of $T$ where $G(T)/G(0) = 1/2$.
Figure~\ref{Fig:9}(b) displays the universal scaling curves
of normalized conductance $G(T)/G(0)$ as a function of $T/T_K$.
For $D=0$, $P=0$ and in the absence of magnetic field, we observe
scaling typical for the underscreened Kondo regime,~\cite{Posazhennikova_PhysRevLett.94.036802}
which has been recently measured experimentally.~\cite{Roch_Phys.Rev.Lett.103/2009}
The temperature dependence of the linear conductance for parameters
where the restoration of the Kondo effect occurs also turns out to be universal, see Fig.~\ref{Fig:9}(b),
however the scaling is completely different from the underscreened Kondo effect.
For parameters where the restoration occurs, the
ground state is two-fold degenerate, i.e. either $\ket{T_+}$ and $\ket{T_0}$
or $\ket{T_0}$ and $\ket{T_-}$ components of the triplet state are degenerate,
therefore one should expect the same scaling as for the spin $S=1/2$ Kondo effect.
~\cite{Kretinin_PhysRevB.84.245316}
Indeed, we compare the universal scaling of the linear conductance for restored Kondo resonances of $S=1$ quantum dot
with the scaling for typical $S=1/2$ Kondo effect and find perfect agreement, see Fig.~\ref{Fig:9}(b).

\section{Summary and conclusions}

By means of numerical renormalization group method, we have
studied transport properties of a magnetic $S=1$ quantum dot
coupled to a ferromagnetic lead in the underscreened Kondo regime.
Due to the coupling of the dot to an external lead, the following two
effective parameters are shown to play important role: the
effective exchange field $B_{\rm eff}$ and the effective
anisotropy constant $D_{\rm eff}$. The interplay of the
corresponding interactions is crucial to understand behavior of
the system transport properties, especially regarding the
evolution (suppression or restoration) of the Kondo effect as a
function of various parameters of the model considered. Using the
second-order perturbation theory, we have derived analytical
formulas for both $B_{\rm eff}$ and $D_{\rm eff}$. It turns out
that the effective anisotropy $D_{\rm eff}$ depends strongly on
$D$ and $D_{\rm eff}\to 0$ as $D\to 0$. Furthermore, $D_{\rm
eff}$  is an even and weakly changing function of the level
position $\varepsilon$, with an extremum at the particle-hole
symmetry point, $\varepsilon = -3U/2$, and does not depend on the
spin polarization of the ferromagnetic lead. The effective
exchange field $B_{\rm eff}$, on the other hand, depends linearly
on lead's spin polarization $P$ and $B_{\rm eff}\to 0$ for $P\to
0$. Furthermore, it is an odd function of level position
$\varepsilon$ and vanishes at the particle-hole symmetry point,
$\varepsilon = -3U/2$. $B_{\rm eff}$ also depends on $D$, although
this dependence is rather weak. We compared the analytical
formulas for $B_{\rm eff}$ and $D_{\rm eff}$ with the NRG data and
found very good agreement.

By performing extensive NRG calculations, we have studied variation
 of the spectral functions with various parameters of the
system. We have shown that the underscreened Kondo effect is
generally suppressed due to the presence of magnetic anisotropy
and exchange field. It can be, however, restored by tuning the
magnetic anisotropy constant $D$. The restoration occurs only for
positive anisotropy, $D>0$, while no restoration takes place when the
magnetic anisotropy is negative, $D<0$. Moreover, the restoration
of the Kondo resonance also occurs as a function of magnetic field
applied along the easy axis. By sweeping the magnetic field, the
Kondo effect can be restored twice in a single sweep. The
restoration always occurs due to the degeneracy between the
components of the triplet state that differ in the spin $S_z$ quantum
number by $1$.

We have also determined the temperature dependence of the linear
conductance for some characteristic parameters, where the
restoration of the Kondo effect occurs. It turned out that the
restored Kondo resonances exhibit a universal scaling as a
function of $T/T_K$ characteristic of spin $S=1/2$ Kondo quantum
dots. This is due to the fact that for parameters where the
restoration of the Kondo effect is possible, the ground state is
two-fold degenerate.

Finally, we would like to emphasize that when considering
spin-resolved transport through nanostructures of spin $S\geq 1$
exhibiting magnetic anisotropy, there are two relevant and
distinct effects that need to be taken into account in order to
fully understand behavior of the system. The first one is the
exchange field induced by ferromagnetic contact, and the second
one is associated with effective (renormalized) magnetic
anisotropy. We also remark that nanoscopic systems for which the
magnetic anisotropy is a generic feature, as the ones discussed in
this paper, present just one possible way of employing
magnetic anisotropy as a key element of novel spintronics devices.
More recently, it has been suggested that spin-anisotropy can also
be generated in \emph{spin-isotropic} systems by spin-dependent
transport of electrons.~\cite{Misiorny_cisa,Baumgartel_Phys.Rev.Lett.107/2011}

\section*{Acknowledgments}

One of us (MM) is  grateful to M. Wegewijs for useful discussions.
This work was supported by the Polish Ministry of Science and Higher Education through a research project in years 2010-2013.
M.M. acknowledges support from
the Foundation for Polish Science and  the Alexander von Humboldt Foundation.
I.W. also acknowledges support from `Iuventus Plus' project for year 2012-2014,
the EU grant No. CIG-303 689, and the Alexander von Humboldt Foundation.



%

\end{document}